\documentclass[prc,twocolumn,aps,nofootinbib, superscriptaddress]{revtex4}

\begin{document}
\date{\today}
\title{{\bf Quantum Monte-Carlo methods and exact treatment of the two-body problem with 
Hartree-Fock Bogoliubov states}}

\author{Denis Lacroix}
\affiliation{National Superconducting Cyclotron Laboratory, Michigan State University, 
East Lansing, Michigan 48824, USA }
\affiliation{
Laboratoire de Physique Corpusculaire,
ENSICAEN and Universit\'e de Caen,IN2P3-CNRS,
Blvd du Mar\'{e}chal Juin,\\
14050 Caen, France}

\begin{abstract}
In this article, we show that the exact two-body problem can be replaced by quantum jumps between 
densities written as  $D=\left| \Psi_a \right> \left< \Psi_b \right|$  
where $\left| \Psi_a \right>$ and $\left| \Psi_b \right>$ are vacuum for different quasi-particles operators.
It is shown that the stochastic process can be written as a Stochastic Time-Dependent Hartree-Fock 
Bogoliubov theory (Stochastic TDHFB) for the generalized density ${\cal R}$ associated to $D$ where 
 ${\cal R}^2 = {\cal R}$ along each stochastic trajectory.   
\end{abstract}
\maketitle

\section{Introduction}
The goal of this article is to prove that the dynamics of fermions interacting through a two-body 
interaction can be transformed into a stochastic process in the Hilbert space of Hartree-Fock Bogoliubov (HFB) states. 
Such a derivation is motivated by recent studies dedicated to the structure of nuclei. In nuclear physics, 
mean-field theories like Hartree-Fock and HFB already provide a good approximation of static and dynamical 
properties \cite{Rin80}. It also turns out that a deep understanding of nuclei requires the introduction 
of correlations beyond mean-field. Large theoretical developments are devoted for instance to Generator Coordinates 
Methods (GCM) \cite{Ben03}. In that case, some collective degrees of freedom are selected and the
correlated ground state is constructed as a superposition of mean-field states (HF or HFB). 
It has been shown that the description of nuclear systems is greatly 
improved if pairing correlations are already accounted for, i.e. if the GCM is performed with 
HFB many-body states.   
Such a method has been successfully applied to nuclear structure when only few degrees of 
freedom are selected. However, applications of GCM techniques when many collective degrees of freedom
are important are still numerically intractable.   

Monte-Carlo techniques appear as an alternative way of treating correlation beyond mean-field.  
Shell model Monte-Carlo theory\cite{Koo97} is an example of such technique.  
Recently, starting from the Hartree-Fock theory, 
new formulations \cite{Car01,Jul02,Lac05} have been proposed that 
combines the advantages of Monte-Carlo methods and mean-field theories.
In that case, the exact evolution of fermionic (or bosonic) systems is replaced by 
an ensemble of stochastic mean-field evolutions. A possible improvement of such theory, which might 
be of great interest in the nuclear context, is to include pairing correlations in the trial set of wave 
functions, i.e. to consider quantum jumps between HFB states instead of HF states. A first step in
this direction as been made in ref. \cite{Mon06} where stochastic dynamics between BCS states were 
introduced. 
In this article, we show that the dynamics of fermions interacting through a general two-body  Hamiltonian 
\begin{eqnarray}
H = \sum_{ij} \left< i \left| T \right| j \right> c^+_i c_j + 
\frac{1}{4} \sum_{ijkl} \left< ij \left| v_{12} \right| lk \right> c^+_i c^+_j c_l c_k 
\end{eqnarray}  
can be mapped into a quantum jump process between HFB states.
Here, $c^+$ operators correspond to creation operators associated to a complete single-particle basis, and $v_{12}$
matrix elements are antisymmetrized. 

In the following we first introduce 
quantities associated to "densities" written as a dyadic of two HFB state vectors, i.e. 
$D= \left| \Psi_a \right> \left< \Psi_b \right|$. The flexibility of stochastic 
methods allows to consider densities with specific helpful 
properties which are precised  in the first part of this work.
Then a TDHFB equation is derived for $D$ when correlations beyond mean-field are neglected. 
Finally, the full stochastic theory that accounts for all two-body effects is derived. 

\section{Preliminary results and Notations}
\label{sec:subclass}

In quantum Monte-Carlo approaches starting from an initial density $D=\left| \Psi \right> \left< \Psi \right|$, the 
exact system evolution is recovered by averaging over an ensemble densities written as a product of two different state vectors
\begin{eqnarray}
D = \left| \Psi_a \right> \left< \Psi_b \right|.
\label{eq:denshfb}
\end{eqnarray}
The use of two different states is at the heart of exact stochastic methods. The advantage of these approaches 
is that states entering in $D$ correspond generally to a specific class of trial wave-function. 
In previous applications, these states have been chosen as Hartree-Fock states\cite{Jul02}. 

Given a specific choice of trial wave-functions, it turns out that stochastic reformulation is generally not unique. This flexibility might
be used for instance to optimize quantum jumps and reduce the number of paths (see for instance \cite{Lac05-2}).
Here we will consider HFB state as trial state vectors and use this flexibility in a different way. 
Indeed, it turns out that the two-body problem can be reformulated as a stochastic process imposing 
additional relations between 
the states $\left| \Psi_a \right>$ and $\left| \Psi_b \right>$ along each path. These additional constraints, 
given below, 
lead to simplified equations and derivations without restricting the exactness of the formulation.

\subsection{Choice of a subclass of densities}

Let us assume that $D$ has the form (\ref{eq:denshfb})
where $\left| \Psi_a \right>$ and $\left| \Psi_b \right>$ can be written as a product of quasi-particle operators, 
i.e. $\left| \Psi_a \right> = \Pi a^+_\alpha \left| 0 \right>$ and $\left| \Psi_b \right> = 
\Pi b^+_\alpha \left| 0 \right>$. $D$ will be referred in the following as a density although 
it does not necessarily meet all required properties to be considered as a density matrix. 
For each quasi-particle operator, two 
sets of single-particle wave-functions, denoted by 
$\left| \alpha_{a,b}  \right>$ and $\left| \bar \alpha_{a,b} \right>$, 
are introduced. They define the transformation between quasi-particle states and a complete set of particle states as
\begin{eqnarray}
\begin{array} {cc}
a_\alpha   = & \sum_{i} c_i \left< \bar \alpha_a \left.  \right| i \right>+ 
\left< i \left.  \right| \alpha_a  \right> c^+_i ,\\
b_\alpha   = & \sum_{i} c_i \left< \bar \alpha_b \left.  \right| i \right>+
 \left< i \left.  \right| \alpha_b  \right> c^+_i .
\end{array}
\label{eq:bogo}
\end{eqnarray}
Note that, we can recover the matrix notations $U,V$ often used in the HFB theory through the 
relation $^{a,b}U_{i\alpha} = \left< \bar \alpha_{a,b} \left.  \right| i \right>$ and 
$^{a,b}V_{i\alpha} = \left< i \left.  \right| \alpha_{a,b}  \right>$. 
   
As usual \cite{Rin80,Bla86}, we introduce vector notations ${\bf a} = \left\{ a, a^+\right\}$, 
${\bf b} = \left\{ b, b^+\right\}$ and ${\bf c} = \left\{ c, c^+ \right\}$. Above linear
transformations can then be written as matrix transformations ${\bf a} = {\cal W}^+_a {\bf c}$ and 
${\bf b} = {\cal W}^+_b {\bf c}$.
In opposite to the standard HFB theory, we do not impose the transformations to be canonical but
instead restrict ourselves to a subclass of quasi-particles and vacuums having two specific properties. 
We first assume that 
\begin{eqnarray}
{\cal W}_a {{\cal W}_b}^+ = {{\cal W}_a}^+ {{\cal W}_b} = 1 ,
\label{eq:prop1}
\end{eqnarray}
which gives the inverse transformations ${\bf c}={\cal W}_b{\bf a} ={\cal W}_a {\bf b}$. 
As a consequence, although the $a$ and $a^+$ operators (respectively the $b$ and $b^+$) 
do not necessarily fulfill fermionic anti-commutation rules, because of (\ref{eq:prop1}) we have 
\begin{eqnarray}
[a_\alpha, b_\beta ]_+ &=& [a^+_\alpha, b^+_\beta ]_+ =  0,~~~
[a_\alpha, b^+_\beta ]_+ = \delta_{\alpha \beta }.
\label{eq:com1}
\end{eqnarray}
The second important assumption is that  $\left| \Psi_a \right>$ and $\left| \Psi_b \right>$ are both vacuum 
for all $a_\alpha$ and $b_\alpha$. As we will see in the following such properties might occur without any simple 
relations between the two sets of annihilation operators.

We introduce the generalized density matrix ${\cal R}_{ab}$ defined as   
\begin{eqnarray}
{\cal R}_{ab} = 
\left( 
\begin{array} {cc}
\left< \Psi_b \left| c^+_i c_j \right| \Psi_a \right>
 & \left< \Psi_b \left| c_i c_j \right| \Psi_a \right> \\
\left< \Psi_b \left| c^+_i c^+_j \right| \Psi_a \right> & \left< \Psi_b \left| c_i c^+_j \right| \Psi_a \right>  
\end{array} \right).
\end{eqnarray}
From the two assumptions, it can be shown that
\begin{eqnarray}
{\cal W}^+_b {\cal R} {\cal W}_a = {\cal L} = 
\left( 
\begin{array} {cc}
0 &0 \\
0 &1 \\
\end{array} \right),
\label{eq:prop2}  
\end{eqnarray}
or equivalently:
\begin{eqnarray}
< a_\alpha b_\beta > &=& < a^+_\alpha b_\beta >= < a^+_\alpha 
b^+_\beta >=0, \nonumber \\
< a_\alpha b^+_\beta >&=&\delta_{\alpha \beta}. \nonumber
\end{eqnarray}
This again can be seen as a generalization of the HFB case and implies 
${\cal R}^2_{ab} ={\cal R}_{ab}$.
In addition, the generalized density ${\cal R}_{ab}$ takes a simplified form 
compared to the one generally obtained for transition densities \cite{Rin80}. Here, we have  
\begin{eqnarray}
{\cal R}_{ab} = 
\left(
\begin{array} {cc}
\rho_{ab} & \kappa_{ab} \\
- \kappa^*_{ab} & 1-{\rho}^T_{ab} 
\end{array}
\right) ,
\end{eqnarray}
where ${\rho}^T_{ab}$ denotes the transposed matrix of $\rho_{ab}$. In the following, 
to simplify notations we will omit the subscript "$_{ab}$".
Different operators matrix elements can be expressed as  
\begin{eqnarray}
\rho &=& \sum_{\alpha}\left| \alpha_a  \right> \left< \alpha_b \right|, \\
1-\rho &=& \sum_{\alpha} \left| \bar \alpha_a  \right> \left< \bar \alpha_b \right|, \\
\kappa &=& \sum_{\alpha}  \left| \alpha_a \bar \alpha_b  \right>= -\sum_{\alpha}  
\left| \bar \alpha_a \alpha_b  \right>,
\end{eqnarray}
with the convention $\kappa_{ij} = \sum_{\alpha}  \left< ij \left.  \right| \alpha_a \bar \alpha_b\right>$.

Finally, we will also use the notation $\left| ^{a,b}W_\alpha \right>$ and $\left| ^{a,b}V_\alpha \right>$ (taken from 
ref. \cite{Bla86}). 
We have in particular 
\begin{eqnarray}
{\cal R} &=& \sum_\alpha \left| {^a}W_\alpha \right> \left<  {^b}W_\alpha \right|, \nonumber \\
1-{\cal R} &=& \sum_\alpha \left| {^a}V_\alpha \right> \left<  {^b}V_\alpha \right|,
\end{eqnarray} 
with $\left<   {^a}W_\beta  \left.  \right|   {^b}W_\alpha  \right>=
\left<   {^a}V_\beta  \left.  \right|   {^b}V_\alpha  \right> =\delta_{\alpha \beta }$ and 
$\left<   {^a}V_\beta  \left.  \right|   {^b}W_\alpha  \right>= 0$. This
completes the different properties associated with the subclass of densities considered here. 


\subsection{Expression of the Hamiltonian and generalized TDHBF equation}

Using the previous properties, the action of the two-body Hamiltonian on the vacuum $\left| \Psi_a \right>$ can be 
recast as  
\begin{eqnarray}
H\left| \Psi_a \right> = \left\{ \left< H \right> + h_L + H^{L}_{res} \right\}\left| \Psi_a \right>,
\label{eq:hpsia}
\end{eqnarray}
where we have used the compact notation 
$\left< H \right> = \left< \Psi_b \left| H \right| \Psi_a \right>$ and where
$h_L$ is a one-body effective Hamiltonian given by
\begin{widetext}
\begin{eqnarray}
h_L = \sum_{\alpha \beta} \left\{ \left< \bar \alpha_b \left| h \right| \beta_a \right> a^+_\alpha b^+_\beta 
+ \frac{1}{2} \Delta_{\bar \alpha_b \bar \beta_b}  a^+_\alpha a^+_\beta
- \frac{1}{2}  \Delta^*_{ \alpha_a \beta_a   }  b^+_\alpha b^+_\beta
\right\}
\label{eq:h1}
\end{eqnarray}
\end{widetext}
$h$ and $\Delta$ correspond respectively to matrix elements 
\begin{eqnarray}
h_{ij}&=& T_{ij} + \left< i\left| Tr_{2} (v_{12}\rho_2) \right| j \right>, \\
\Delta_{ij} &=& \frac{1}{2} \sum_{kl} \left< ij \left| v_{12} \right| kl \right> \kappa_{kl},
\end{eqnarray}  
which will be called mean-field and pairing field in analogy to HFB theory.
Note that expression (\ref{eq:h1}) differs from the one generally obtained in HFB using the Wick theorem because of
the 
coexistence of two sets of quasi-particle operators. 
Starting from (\ref{eq:h1}), the effective Hamiltonian can be recast as
\begin{eqnarray}
h_L = \frac{1}{2}(
\begin{array} {cc}
c^+ & ~~c 
\end{array})
(1-{\cal R}) {\cal H} {\cal R} 
\left(
\begin{array} {c}
c \\ 
c^+ 
\end{array}
\right) ,
\label{eq:h1rr}
\end{eqnarray}
where ${\cal H}$ stands for the generalized HFB Hamiltonian \cite{Rin80,Bla86}:
\begin{eqnarray}
{\cal H} = \left( 
\begin{array} {ccc}
h & \Delta \\
-\Delta^* & -h^T
\end{array} \right).
\end{eqnarray}
We will see that expression (\ref{eq:h1rr}) is central for further developments. 
The last term of equation (\ref{eq:hpsia}), called hereafter
residual Hamiltonian reads  
\begin{eqnarray}
H^{L}_{res} = \frac{1}{4} \sum_{\alpha \beta \gamma \delta} \left< \bar \alpha_b \bar \beta_b \left| V_{12} \right| \delta_a \gamma_a \right>
a^+_\alpha a^+_\beta b^+_\gamma b^+_\delta.
\end{eqnarray}
Performing a similar decomposition of $\left< \Psi_b \right| H$ leads to
\begin{eqnarray}
\left< \Psi_b \right| H = \left< \Psi_b \right| \left\{ \left< H \right> + h_R + H^{R}_{res} \right\},
\end{eqnarray}
with 
\begin{eqnarray}
h_R = 
\frac{1}{2}(
\begin{array} {cc}
c^+ & ~~c 
\end{array})
{\cal R} {\cal H} (1-{\cal R})
\left(
\begin{array} {c}
c \\ 
c^+ 
\end{array}
\right) ,
\end{eqnarray}
while
\begin{eqnarray}
H^{R}_{res} = \frac{1}{4} \sum_{\alpha \beta \gamma \delta} \left< \alpha_b \beta_b \left| v_{12} \right| \bar \delta_a \bar 
\gamma_a \right> a_\alpha a_\beta b_\gamma b_\delta.
\end{eqnarray}

\subsection{Evolution of the generalized density ${\cal R}$}
Starting from the initial density (\ref{eq:denshfb}), 
the evolution of the system 
is considered assuming first that the effect of the residual interaction can be neglected,. 
After one time-step, due to the one-body nature of $h_L$, the state 
$\left| \Psi_a + d\Psi_a \right> = e^{\frac{dt}{i\hbar}h_L} \left| \Psi_a \right>$
is a  vacuum for the new quasi-particles 
$a'_\alpha = a_\alpha + d a_\alpha = 
e^{\frac{dt}{i\hbar}h_L} a_\alpha e^{-\frac{dt}{i\hbar}h_L}$. Similarly, 
$\left< \Psi_b + d\Psi_b \right|= \left< \Psi_b \right| e^{-\frac{dt}{i\hbar}h_R}$ 
is a vacuum for the new quasi-particles 
${b'_\alpha}^+ = b^+_\alpha + d b^+_\alpha = 
e^{\frac{dt}{i\hbar}h_R} b^+_\alpha e^{-\frac{dt}{i\hbar}h_R}$. 

Since the residual interaction is neglected, all 
the information on the system is contained in the evolution of ${\cal R}$. 
From standard rules of creation-annihilation operator 
transformations \cite{Blo62,Bal69,Bla86}, we obtain:
\begin{eqnarray}
\left[h_L, {\bf c}\right] &=& -(1-{\cal R}) {\cal H} {\cal R}{\bf c}, \\
\left[h_R, {\bf c}\right] &=& -{\cal R}{\cal H} (1-{\cal R}) {\bf c}.
\label{eq:com}
\end{eqnarray}
With the help of the above anti-commutation relationships, we can express $e^{-\frac{dt}{i\hbar}h_L} {\bf c}
e^{+\frac{dt}{i\hbar}h_L}$ and $e^{\frac{dt}{i\hbar}h_R} {\bf c} 
e^{-\frac{dt}{i\hbar}h_R}$, and deduce the evolution of ${\cal R}$.
Using the fact that initially ${\cal R}^2 = {\cal R}$ and 
$\left( 1 - {\cal R} \right){\cal R} = 0$, we end with
\begin{eqnarray}
i \hbar \frac{d {\cal R}}{dt} &=& (1-{\cal R}) {\cal H} {\cal R}  - {\cal R}{\cal H} (1-{\cal R}) \nonumber \\
&=& \left[{\cal H}, {\cal R}  \right],
\end{eqnarray}
which is nothing but a TDHFB equation generalized to densities given by eq. (\ref{eq:denshfb}). Without going into further 
details, it can be shown that the density ${\cal R}$ 
fulfills all properties listed above and thus remains in 
the subclass of densities  previously described. Therefore, the considerations made 
for single time step can be extended to the long-time dynamics.    

\section{Introduction of Quantum Monte-Carlo methods}
In the previous section, we have introduced general properties of densities given by eq. (\ref{eq:denshfb})
which will be helpful for the forthcoming discussion. In addition, we have shown that the dynamics reduces 
to a TDHFB-like equation when the residual interaction is neglected. 
The aim of this section is to show that the residual interaction can be treated by introducing 
stochastic processes between densities described in section \ref{sec:subclass}.

\subsection{Separable residual interaction}
In 
the following discussion, we concentrate first on the evolution of $\left| \Psi_a \right>$, keeping in
mind that everything can be transposed to $\left< \Psi_b \right|$.
Following the Stochastic mean-field approach \cite{Car01,Jul02,Lac05}, we consider that 
the residual part of the interaction can be written as a sum of of separable interactions in the particle-hole channel: 
\begin{eqnarray}
\left<\bar \alpha_a \bar \beta_a  \left| v_{12} \right| 
\delta_b \gamma_b  \right> = - \sum_{m} \left<\bar \alpha_a  \left| O_m \right| 
\delta_b \right> \left< \bar \beta_a  \left| O_m \right| 
\gamma_b  \right> ,
\label{eq:sepph}
\end{eqnarray}  
where $O_m$ corresponds to a set of single-particle operators.
Using relation (\ref{eq:sepph}), the residual interaction $H^{L}_{res}$ can be recast as  
\begin{eqnarray}
H^{L}_{res} \left| \Psi_a \right>= \frac{1}{4} \sum_{m} B^{ph}_m  B^{ph}_m  \left| \Psi_a \right>,
\label{eq:hlquare}
\end{eqnarray}
where the set of one-body operators $B_m$ are given by
\begin{eqnarray}
B^{ph}_m = \sum_{ \alpha \beta } \left< \bar \alpha_a \left| O_m \right| \beta_b \right> 
a^+_\alpha b^+_\beta .
\end{eqnarray}
Guided by previous section, we write it as 
\begin{eqnarray}
B^{ph}_m = \frac{1}{2} \left( 
\begin{array} {cc}
c & c^+  
\end{array}
\right) (1-{\cal R}) {\cal B}^{ph}_m {\cal R}
\left( 
\begin{array} {c}
c^+ \\
c  
\end{array}
\right) ,
\label{eq:bph}
\end{eqnarray} 
where we have introduced the matrix ${\cal B}^{ph}_m$:
\begin{eqnarray}
{\cal B}^{ph}_m =
\left(
\begin{array} {ccc}
O_m & 0 \\
0 & -O_m^{T} 
\end{array} 
\right).
\end{eqnarray}
Once $H^L_{res}$ is written as (\ref{eq:hlquare}), the introduction of stochastic process is rather 
straightforward. Introducing a set of stochastic  variables $d \xi^{L}_{m}$ 
(which follow Ito rules of stochastic calculus \cite{Gar85}) with mean values equal to zero and variances 
satisfying
\begin{eqnarray}
\overline{d \xi^{L,(n)}_{m} d \xi^{L,(n)}_{m'}} = \delta_{m m'} \frac{d t} {2 i\hbar}.
\end{eqnarray} 
Here the $^{(n)}$ exponent stands for a specific realization of the stochastic process. In the following, 
it will sometimes be omitted to simplify notations.
The evolution of $\left| \Psi_a \right>$ associated to $H$ can then be written as an average
over stochastic evolutions in the Hilbert space of HFB state vectors: 
\begin{eqnarray}
e^{\frac{dt}{i \hbar} H}\left| \Psi_a  \right> &=& e^{\frac{d t}{i \hbar} \left<H \right>}~\overline{
e^{\frac{d t}{i \hbar} h_L + \sum_m d\xi^{L}_m B^{ph}_m} \left| \Psi_a  \right> } \\
&=& \overline{\left|\Psi_a^{(n)}(t+dt) \right> }.
\end{eqnarray}  
In this equation, same conventions as in ref. \cite{Lac05,Bre02} are used and $\left| \Psi^{(n)}_a (t+dt) \right>$
correspond to different vacuum states. Introducing the notation
\begin{eqnarray}
dS^{L}_{ph} = \frac{d t}{i \hbar} h_L + \sum_m d\xi^{L}_m B^{ph}_m,
\end{eqnarray} 
according to the Thouless theorem, each $\left| \Psi^{(n)}_a \right>$ is a vacuum for the quasi-particle 
$a'_\alpha = e^{dS^{L}_{ph}} a_\alpha e^{-dS^{L}_{ph}}$.

One can finally note that the stochastic evolution of $\left| \Psi_a \right>$ should be completed 
by an equivalent stochastic evolution for $\left< \Psi_b \right|$. The 
associated propagator and stochastic variables are respectively denoted by 
$dS^R_{ph}$ and $d\xi^R_m$. 
These variables should be taken statistically independent 
of $d\xi^L_m$ to properly account for the exact dynamics of $D$.  
More explicitly, we have 
\begin{eqnarray}
dS^R_{ph} = -\frac{d t}{i \hbar} h_R + \sum_m d\xi^{R}_m {B^{ph}_m},
\end{eqnarray}
where 
\begin{eqnarray}
B^{ph}_m = \frac{1}{2} \left( 
\begin{array} {cc}
c & c^+  
\end{array}
\right)  {\cal R}{\cal B}^{ph}_m (1-{\cal R})
\left( 
\begin{array} {c}
c^+ \\
c  
\end{array}
\right),
\label{eq:bph}
\end{eqnarray} 
and $\overline{d\xi^{R}_md\xi^{R}_{m'}} = -\delta_{mm'}\frac{dt}{i\hbar}$.
   
\section{Nature of the stochastic process}

Similarly to the previous case, where the residual interaction was neglected, we do expect that
along each path, the stochastic evolution of the system reduces to the stochastic evolution 
of ${\cal R}$. The explicit form of the stochastic evolution of ${\cal R}$ can now be obtained 
using the commutation relationship \cite{Blo62,Bal69,Bla86}:
\begin{eqnarray}
\left[e^{dS^L_{pp}}, {\bf c}  \right] &=& e^{-(1-{\cal R}) 
\left[\frac{dt}{i\hbar}{\cal H} + {\cal B}^L \right]{\cal R}} {\bf c} \nonumber \\
&=& {\bf c}-(1-{\cal R}) \left[\frac{dt}{i\hbar}{\cal H} + {\cal B}^L \right]{\cal R} {\bf c},
\label{eq:commut1}
\end{eqnarray}
while 
\begin{eqnarray}
\left[e^{dS^R_{pp}}, {\bf c}  \right] &=& e^{-{\cal R}
\left[-\frac{dt}{i\hbar}{\cal H} + {\cal B}^R \right](1-{\cal R})}  {\bf c} \nonumber \\
&=& {\bf c}- {\cal R} \left[-\frac{dt}{i\hbar}{\cal H} + {\cal B}^R \right](1-{\cal R}){\bf c},
\label{eq:commut2}
\end{eqnarray}
where ${\cal B}^L$ and ${\cal B}^R$ stand for
\begin{eqnarray}
{\cal B}^{L/R} = \sum_m d\xi^{L/R}_m {\cal B}^{ph}_m. 
\end{eqnarray} 
Note that equations (\ref{eq:commut1}) and (\ref{eq:commut2}) are exact 
thanks to the $(1-{\cal R})$ term. Similarly, as in previous section 
and using expression (\ref{eq:commut1}) and (\ref{eq:commut2}), one gets the stochastic evolution of 
${\cal R}$:
\begin{eqnarray}
d{\cal R} =\frac{dt}{i \hbar} \left[ {\cal H}, {\cal R} \right]+(1-{\cal R}){\cal B}^{L}{\cal R}  
+ {\cal R}{\cal B}^{R}(1-{\cal R})  .
\label{eq:STDHFB}
\end{eqnarray}  
Such a stochastic process, called hereafter Stochastic TDHFB, is similar to the 
Stochastic mean-field dynamics \cite{Jul06} except that 
the mean-field and normal densities are now replaced respectively by the HFB Hamiltonian ${\cal H}$
and density ${\cal R}$. Starting from ${\cal R} = \sum_\alpha \left| {^a}W_\alpha \right> 
\left<  {^b}W_\alpha \right|$, evolution of ${\cal R}$ can be replaced by the set of equations
\begin{eqnarray}
\left\{
\begin{array} {ccc}
\left| d{^a}W_\alpha \right> &=& \left\{ \frac{dt}{i \hbar}  {\cal H}
+(1-{\cal R}){\cal B}^{L}  \right\} \left| {^a}W_\alpha \right> \\
&& \\
\left< d{^b}W_\alpha \right| &=&\left<  {^b}W_\alpha \right| 
\left\{  -\frac{dt}{i \hbar} {\cal H} + {\cal B}^{R}(1-{\cal R}) \right\}
\end{array}
\right.
.
\label{eq:dw} ,
\end{eqnarray}
Above expressions show that if initially fullfilled, 
the property $\left<   {^a}W_\beta  \left.  \right|   {^b}W_\alpha  \right>
= \delta_{\alpha \beta }$ is true all along each stochastic path. Again, 
it can be shown that all the properties of the class of densities considered in section \ref{sec:subclass} 
are preserved during the stochastic evolution (\ref{eq:STDHFB}). Therefore, we only have to initiate 
the quantum jump process with a density which  satisfies 
the properties described in section \ref{sec:subclass}. This is the case if   
we start from an initial HFB density $D=\left| \Psi \right> \left< \Psi \right|$, which is the 
most convenient in practice.  
Note finally that the explicit form of the quasi-particle 
evolution can directly be obtained from eq. (\ref{eq:dw}) while the stochastic evolution
of $\rho$ and $\kappa$ can be deduced from (\ref{eq:STDHFB}). 

\subsection{Alternative form and $pp-hh$ separable interaction}
In the previous section, we have developed quantum diffusion processes between HFB states 
assuming expression (\ref{eq:sepph}). However, 
recent studies in nuclear structure \cite{Dug04} support separable interactions  in the particle-particle and 
hole-hole channels. For completness, we introduce a set of one-body operators $G_m$ and assume 
that the residual interaction now reads 
\begin{eqnarray}
\left<\bar \alpha_a \bar \beta_a  \left| v_{12} \right| 
\delta_b \gamma_b \right> = - \sum_{m} \left<\bar \alpha_a  \left| G_m \right| 
\bar \beta_b \right> \left< \delta_a  \left| G_m \right| 
 \gamma_b  \right>^*,
\label{eq:seppp} 
\end{eqnarray}
where $G_m$ should be a skew matrix (i.e. $G^T_m = -G_m$) to respect the antisymmetrization of $v_{12}$. 
The formulation of the two-body problem can equivalently be done starting from eq. (\ref{eq:seppp}). 
The final result is that ${\cal R}$ still obeys a stochastic equation with similar form as (\ref{eq:STDHFB}), where
${\cal B}^{L/R}$ now reads 
\begin{eqnarray}
{\cal B}^{L/R} = \sum_m d\eta^{L/R}_m {\cal B}^{pp}_m + i \left( d\eta^{L/R}_m \right)^* {\cal B}^{hh}_m,
\end{eqnarray}
and $d\eta_m$ corresponds to stochastic variables with mean value zero and
\begin{eqnarray}
\overline{d \eta^{L}_{m} \left( d \eta^{L}_{m'} \right)^*} 
&=& \delta_{m m'} \frac{d t} {2 \hbar},
\end{eqnarray}
while all other second moments are equal to zero. 
${\cal B}^{pp}_m$ and ${\cal B}^{hh}_m$ are given by
\begin{eqnarray}
{\cal B}^{pp}_m =
\left(
\begin{array} {ccc}
0 & G_m \\
0 & 0 
\end{array} 
\right),~~~~~~ 
{\cal B}^{hh}_m =
\left(
\begin{array} {ccc}
0 &  0 \\
-G^*_m & 0 
\end{array} 
\right).
\end{eqnarray}

\section{Conclusions}

In this work, we have shown that the exact dynamics of interacting fermions 
can be replaced by a Monte-Carlo method in the Hilbert space of Hartree-Fock Bogoliubov 
states. In order to prove this reformulation, we have used an intermediate 
result, considering densities $D=\left| \Psi_a \right> \left< \Psi_b \right|$ with specific 
properties. Neglecting the residual interaction, the evolution of $D$ leads to a TDHFB
equation for ${\cal R}$. We then have proven that the introduction of correlations beyond the 
mean-field picture can be 
replaced by a Stochastic TDHFB equation for ${\cal R}$, generalizing the stochastic mean-field approach \cite{Jul02}. 
It should be noted that the reformulation is not unique and the selection of a sub-class of $D$ is not absolutely necessary. 
However, in that case derivations and stochastic equations are more complicated.

The stochastic theory presented here is not restricted to dynamical problem and could also be useful
to study static properties of interacting systems \cite{Jul02}. In that case, real time propagation is 
replaced by imaginary time evolution. Monte-Carlo methods has the advantage of not requiring an a priori 
knowledge of the relevant collective 
degrees of freedom and can eventually be used as an alternative to GCM. It should however be noted that Monte-Carlo
methods still require large numerical efforts. Work is actually in progress to combine advantages of GCM 
and Monte-Carlo techniques.     
      
Finally, we would like to mention that the above theory gives an indirect proof of the fact that 
densities described in section \ref{sec:subclass} form an over-complete set of densities to treat 
the two-body problem. This might be of great interest even for non stochastic methods which  
treat correlations beyond mean-field.

{\bf ACKNOWLEDGMENTS}

The author is grateful to Thomas Duguet and Vincent Rotival for
the careful reading of the manuscript and to Olivier Juillet for
discussions during this work.


\begin{references}
\bibitem{Rin80}  {P. Ring and P. Schuck, {\it The Nuclear Many-Body Problem,
Spring-Verlag}, New-York (1980).}
\bibitem{Ben03} M. Bender, P.-H. Heenen, and P.-G. Reinhard, Rev. Mod. Phys. 75, 121 (2003).
\bibitem{Koo97}  S.E.Koonin, D.J.Dean, K.Langanke, Ann.Rev.Nucl.Part.Sci. 
{\bf 47} (1997) 463. 
\bibitem{Car01} I. Carusotto, Y. Castin and J. Dalibard, Phys. Rev. A63
(2001) 023606.
\bibitem{Jul02}  O. Juillet and Ph. Chomaz, Phys. Rev. Lett. {\bf 88} (2002)
142503.
\bibitem{Lac05} D. Lacroix, Phys. Rev. {\bf C71}, 064322 (2005).
\bibitem{Mon06} {A. Montina and Y. Castin, Phys. Rev. {\bf A73}, 013618 (2006).}


\bibitem{Lac05-2} D. Lacroix, Phys. Rev. A 72, 013805 (2005). 

\bibitem{Bla86}  J.P. Blaizot and G. Ripka, Quantum Theory of Finite
Systems, (MIT Press, Cambridge, Massachusetts, 1986).

\bibitem{Blo62} {C. Bloch, lecture notes on Nuclear Many-Body problem, Bombay, (1962)}
\bibitem{Bal69} R. Balian and E. Brezin, Nuovo Cimento {\bf B64}, 37 (1969).

\bibitem{Gar85}  W. Gardiner, ''Handbook of Stochastic Methods'',
Springer-Verlag, (1985).

\bibitem{Bre02} H.P. Breuer and F. Petruccione, \textit{The Theory of Open
Quantum Systems} (Oxford University Press, Oxford, 2002).


\bibitem{Jul06} {O. Juillet, {\it in preparation}.}

\bibitem{Dug04}  T. Duguet, Phys. Rev. C 69, 054317 (2004). 


\end{references}
\end{document}